# Citizen Centered Climate Intelligence: Operationalizing Open Tree Data for Urban Cooling and Eco-Routing in Indian Cities

Kaushik Ravi[1] and Andreas Brück[2]

[1]Department of Civil Engineering, National Institute of Technology Tiruchirappalli

[2]Institute for Urban and Regional Planning, Technische Universität Berlin

**Author Note**

Kaushik Ravi  https://orcid.org/0009-0009-5015-7464

Andreas Brück  https://orcid.org/0000-0002-5286-3700

The authors have no known conflicts of interest to disclose.

This research did not receive any specific grant from funding agencies in the public, commercial, or not-for-profit sectors.

The authors used an AI-based tool (Google AI Studio) to improve the language and readability of the manuscript. All content was subsequently reviewed and edited by the authors, who take full responsibility for the final text.

Correspondence concerning this article should be addressed to Kaushik Ravi.

Email: official.kaushik.r@gmail.com

**Abstract**

Urban climate resilience requires more than high-resolution data; it demands systems that embed data collection, interpretation, and action within the daily lives of citizens. This chapter presents a scalable, citizen-centric framework that reimagines environmental infrastructure through participatory sensing, open analytics, and prescriptive urban planning tools. Applied in Pune, India, the framework comprises three interlinked modules: (1) a smartphone-based measurement toolkit enhanced by AI segmentation to extract tree height, canopy diameter, and trunk girth; (2) a percentile-based model using satellite-derived Land Surface Temperature to calculate localized cooling through two new metrics, Cooling Efficacy and Ambient Heat Relief; and (3) an eco-routing engine that guides mobility using a Static Environmental Quality score, based on tree density, species diversity, and cumulative carbon sequestration. Together, these modules form a closed feedback loop where citizens generate actionable data and benefit from personalized, sustainable interventions. This framework transforms open data from a passive repository into an active platform for shared governance and environmental equity. In the face of growing ecological inequality and data centralization, this chapter presents a replicable model for citizen-driven urban intelligence, reframing planning as a co-produced, climate-resilient, and radically local practice.

# Citizen Centered Climate Intelligence: Operationalizing Open Tree Data for Urban Cooling and Eco-Routing in Indian Cities

## Introduction

A quiet revolution is underway in how we understand, manage, and navigate our cities. At the intersection of data, cities, and citizens, the rise of open data is creating unprecedented opportunities to revolutionize urban and transport planning. This shift challenges the traditional, top-down models of city governance, which often fail to address the hyperlocal complexities of urban life. Our chapter explores the frontier of this revolution, demonstrating how citizen-generated open data can be operationalized to build more climate-resilient and equitable urban environments.

This challenge is particularly acute in the context of the climate crisis, which manifests uniquely at the street level. In India, for instance, cities are warming at an alarming rate of 0.53°C per decade, nearly double that of surrounding rural areas, posing a direct threat to public health and economic stability (Mongabay-India, 2024). Conventional, expert-led planning often operates at a scale mismatched with the lived realities of residents (Bulkeley & Castán Broto, 2013), creating a critical gap between macro-level policy and the micro-level conditions where climate impacts are most acutely felt. Bridging this gap requires new paradigms that not only generate granular data but also empower citizens to become active agents in shaping their own communities.

This chapter advocates for a new paradigm centred on citizen-led data generation and action. We propose an integrated framework that harnesses the synergistic potential of Citizen Science and Open Data to address the scale mismatch in urban climate governance. Citizen science, defined as the active participation of the public in scientific research, enables the generation of data at a spatiotemporal resolution previously unattainable (Irwin, 1995). When coupled with publicly available Open Data, it provides a verifiable and transparent foundation for building trust, fostering innovation, and empowering communities (Open Knowledge Foundation, 2015).

## An Integrated Framework for Urban Analysis

This framework is operationalized through three interlocking, technologically enabled modules that form a complete, end-to-end data ecosystem:

- A Democratized Measurement Toolkit: A suite of methods cantered on smartphone photogrammetry that empowers non-experts to accurately measure key urban tree metrics (e.g., DBH, height), transforming citizens into active data producers.

- An Interactive Analytics Dashboard: A visualization platform that synthesizes tree census data, citizen-contributed measurements, and remote sensing data into a coherent and explorable interface for planners to analyse urban ecosystem services and simulate greening interventions.
- A Holistic Environmental Routing Engine: An advanced routing algorithm that moves beyond simple time-and-distance optimization to incorporate environmental variables, providing users with actionable, sustainable mobility choices.

To operationalize this, we present an integrated system of three interlocking modules: a Democratized Measurement Toolkit for citizen-led data generation; an Interactive Analytics Dashboard for planners; and a Holistic Eco-Routing Engine for individual mobility choices. This chapter details the design and implementation of this end-to-end framework, demonstrating a replicable model that transforms citizens from passive subjects into active agents of urban climate action.

## Theoretical Framework

Our research is situated at the intersection of three dynamic fields: urban ecosystem modelling, high-resolution environmental sensing, and sustainable mobility systems. A critical analysis of the literature reveals a persistent fragmentation between these domains. This review constructs the theoretical argument for our integrated framework by examining the limitations of existing models, exploring the potential of new sensing paradigms, and identifying the gap in translating new data into meaningful action.

### Modelling Urban Ecosystems

The scientific bedrock for quantifying terrestrial carbon stocks is the allometric equation. The pantropical models developed by Chave et al. (2014) are a landmark in this field, offering a robust method for estimating tree biomass from core metrics such as diameter at breast height (DBH), height, and wood density. However, the direct application of these forest-centric models to urban ecosystems poses significant challenges. Urban landscapes are not natural forests; they are complex mosaics of managed and highly stressed environments. Research in the Indian context has quantified this discrepancy, revealing that the carbon sequestration of roadside trees can be 18-22% lower than that of park trees of the same species (Singh et al., 2022). This highlights a fundamental limitation: general allometric models are often blind to the hyperlocal variations in urban typology that significantly influence tree physiology.

While our framework currently employs these established pantropical equations as the best available starting point, its core architectural purpose is to address this very data gap. By creating a mechanism for collecting vast, granular, and location-specific data, our system builds the essential repository required to calibrate existing models or develop new, context-aware urban allometric equations in the future. This moves beyond a static application of a generic model by creating a dynamic feedback loop where citizen-generated data can eventually lead to more refined and accurate urban-centric science.

**Sensing the City at a Human Scale**

Concurrent with advances in modelling has been a revolution in the resolution of environmental data. At the macro-scale, satellite remote sensing platforms like Landsat have been instrumental in mapping regional phenomena such as Land Surface Temperature (LST) and the Urban Heat Island effect (Almeida et al., 2022). Yet, this top-down perspective often misses the street-canyon microclimates where human life unfolds. To bridge this gap, a move towards ground-level, or "human-scale," sensing has emerged, from using Google Street View to quantify shade equity (Li & Ratti, 2018) to deploying fixed sensor networks (Pathak, 2023).

It is here that we position Citizen Science not merely as a data collection method, but as a transformative research paradigm (Irwin, 1995; Cohn, 2008). This approach recasts the public as an active participant in research, which, from a geoinformatics perspective, creates a network of "citizens as sensors" (Goodchild, 2007). This concept of Volunteered Geographic Information (VGI) offers a pathway to generate datasets of a scale and granularity that are otherwise logistically infeasible, fostering a co-creation process between researchers and the public.

**From Citizen Sensing to Citizen-Centred Intelligence**

The term "citizen-centred" is foundational to our framework and warrants a precise definition, distinguishing it from related concepts. Unlike "user-centred design," which primarily focuses on the usability and experience of a tool for an individual, our "citizen-centred" approach considers the collective and civic implications of data generation and use, situating our work within the critical discourse on the 'data revolution' and its impact on cities (Kitchin, 2014). Furthermore, while not entirely "citizen-led", in that the initial framework was designed by researchers, it is "citizen-centred" because it is architected to empower citizens by translating their individual contributions into collective intelligence and actionable insights. This empowerment materializes through three distinct roles for the citizen:

1. Citizen as Producer: In this role, citizens act as a distributed sensor network, using the Democratized Measurement Toolkit to generate hyperlocal environmental data at a scale unattainable by official surveys. This directly addresses the concept of "citizens as sensors" (Goodchild, 2007) and enriches the urban dataset with granular, on-the-ground knowledge.
2. Citizen as User: As the direct beneficiary of the data, the citizen uses the Eco-Routing Engine to make informed mobility choices. This creates a tangible, personal benefit, closing the feedback loop between data contribution and personal utility, and transforming citizens from passive data subjects into active participants in a more sustainable transport system.
3. Citizen as Advocate: The framework's most transformative potential lies in positioning the citizen as an advocate. By providing transparent, data-driven tools and verifiable metrics on ecosystem services, we equip communities with a new form of evidence-based agency. This approach aligns with the upper rungs of (Arnstein's 1969) 'Ladder of Citizen Participation,' aiming to move citizens from mere tokenism towards a genuine partnership in the planning process, a challenge that persists even in the age of digital 'smart cities' (Cardullo & Kitchin, 2019).

**Integrating Data into Sustainable Mobility**

The final piece of our theoretical framework addresses the critical question: how can this new, high-resolution environmental data be translated into meaningful action? The domain of transport and mobility provides a powerful answer. Conventional routing algorithms, optimizing solely for time or distance, are increasingly recognized as contributors to negative environmental externalities (Barkenbus, 2010). The emerging field of eco-routing offers an alternative, yet early innovations often focused on human-perceived qualities. For example, groundbreaking work has demonstrated the possibility of routing for subjective experiences, such as beauty, quietness, and happiness, by using crowdsourced perceptual data (Quercia et al., 2014).

While this marked a critical shift from pure efficiency to human experience, a significant gap persists: the lack of routing systems capable of integrating objective, dynamic, granular, and positively-weighted ecosystem services. Existing systems may route a user through a park, but they typically cannot differentiate between a route lined with a monoculture of young, low-sequestration trees and one bordered by a diverse canopy of mature, high-sequestration giants. Our research directly addresses this disconnect by arguing that the next leap forward requires a holistic framework that integrates better models, participatory sensing, and smarter, value-driven algorithms into a single, functional feedback loop.

## Methodology: An End-to-End Citizen-Centric Pipeline

Our framework operationalizes the synthesis of citizen-led measurements, municipal open data, and remote sensing imagery through a scalable, end-to-end pipeline. The process transforms these raw data streams into high-value, prescriptive analytics for urban climate action. In the spirit of transparency and reproducibility, the source code for the prototype is available for review https://github.com/Kaushik-Ravi/citizen-climate-intelligence.

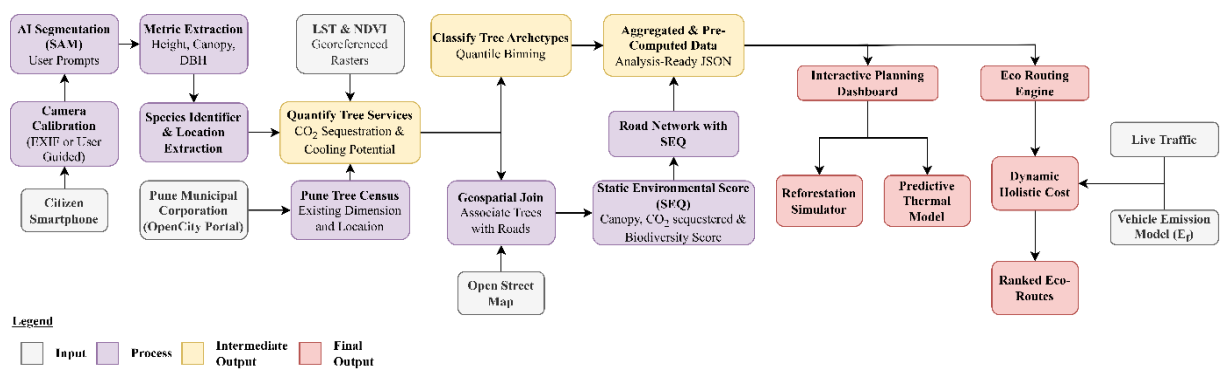

**Figure 1**

*Conceptual Architecture of the Integrated Framework*

*Note.* The flowchart illustrates the synergistic relationship between the three core modules and their data flows, moving from data generation (left) to urban-scale analysis (center) and individual-level action (right).

### Module 1: The Democratized Measurement Toolkit

The foundational data-gathering component of our system is a novel framework for interactive photogrammetric dendrometry, designed to enable precise data acquisition using uncalibrated, commodity smartphones. This approach aligns with an emerging field of research validating the use of low-cost, close-range photogrammetry for forest inventory (Mokroš et al., 2018).

The model is grounded in the geometric principles of the pinhole camera, which establishes a relationship of similar triangles between the 3D object space and the 2D image plane. The central objective is to derive a $Scale\ Factor$, $S$, which defines the real-world distance subtended by a single pixel. This is achieved by first determining the camera's intrinsic $Camera\ Constant$, $C$, which is the unitless ratio of the camera's sensor width ($W_s$) to its focal length ($f$). The real-world scene width ($W_{scene}$) captured by the camera at a given distance to an object ($D_o$) can then be expressed as:

$$W_{\text{scene}} = D_o \cdot C \text{ where } C = \frac{W_s}{f} \quad (1)$$

The *Scale Factor*, $S$, which translates pixel measurements into metric units, is then calculated by dividing the real-world scene width by the image width in pixels ($P_{width}$).

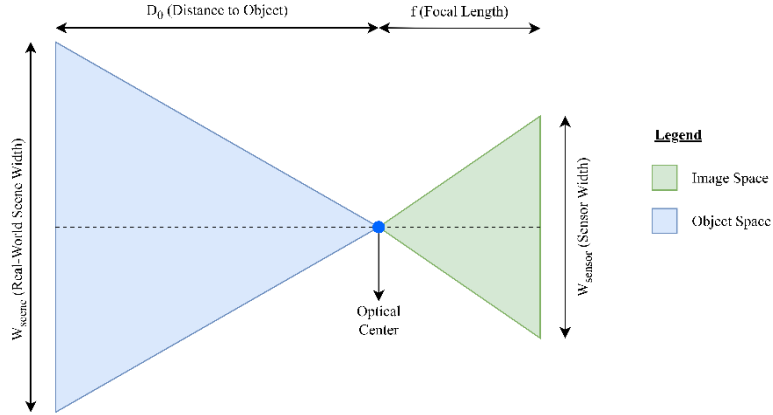

**Figure 2**

*The Pinhole Camera Model*

*Note.* The diagram illustrates the principle of similar triangles that underpins the photogrammetric model, establishing the geometric relationship between the 3D object space and the 2D image plane.

A key challenge in using commodity smartphones is that intrinsic camera parameters are sometimes inaccessible or unreliable. Our framework addresses this via a dual-pathway system for robustly determining the *Camera Constant*, $C$:

1. Pathway 1: EXIF-Based Estimation. The primary pathway leverages standardized EXIF metadata from an image to compute $C$ automatically.
2. Pathway 2: User-Guided Calibration. In the absence of reliable EXIF data, a secondary pathway facilitates a user-guided in-situ calibration, a process that reverse-engineers the *Camera Constant* through the observation of a reference object of known size at a known distance.

To maximize accuracy and user agency, the toolkit then employs a multi-modal measurement framework. The primary modality is designed for features with complex shapes, such as tree canopies, and employs the Segment Anything Model (SAM), a powerful foundation model for image segmentation (Kirillov et al., 2023), for AI-assisted segmentation.

The toolkit is engineered to address the practical challenges and potential validity issues inherent in field-based citizen science. We explicitly acknowledge that the validity of 2D photogrammetry is contingent upon the camera being held roughly orthogonal to the subject.

Indeed, empirical testing of our underlying photogrammetric model under controlled conditions; optimal lighting, distances of 5-6 meters, and near-orthogonal camera angles, confirms a high degree of precision, yielding measurement accuracy of ±2 cm. This quantitative validation under ideal conditions provides a robust baseline for the toolkit's potential. However, our design philosophy assumes that real-world conditions are rarely ideal. A prime example of this is the Smart DBH feature. As documented in our development process, early automation attempts struggled with scenarios like trunk occlusion. The Smart DBH feature is a direct result of this, empowering the user to select any visible portion of the trunk, thereby ensuring a valid measurement can be taken even in challenging field conditions. This design philosophy, balancing a validated, high-precision model with pragmatic, user-driven heuristics, is central to the toolkit's practicality and its potential for successful deployment.

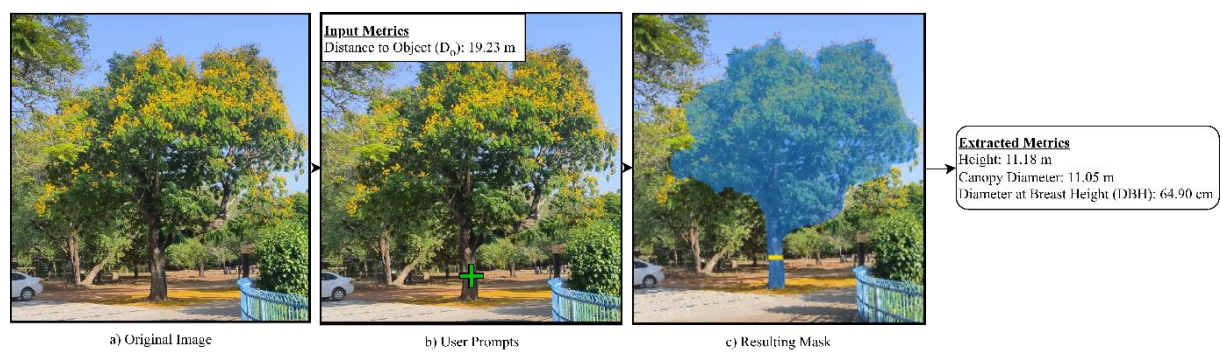

**Figure 3**

*Tree structure extraction via SAM-assisted image segmentation.*

*Note*. From a single user image and prompt, the model estimates tree height, canopy width, and DBH. This pipeline supports non-expert measurement of tree parameters using standard smartphones.

**Module 2: The Analytics Engine for Urban Planning**

This module processes multi-source data streams to produce robust metrics of ecosystem services and generate prescriptive, actionable tools for urban planners.

We quantify two primary ecosystem services. The first is $CO_2$ sequestration. This process begins by estimating the Above-Ground Biomass (AGB) for each tree using the Pan-Tropical Allometric Equations (Chave et al., 2014), which utilize the tree's dimensions and species-specific wood density. This AGB is then converted to a final $CO_2$ equivalent through a standardized, multi-step process. First, Total Biomass is calculated by accounting for the root-to-shoot ratio, using a factor of 0.26, which is a value appropriate for the tropical dry forest context relevant to India (Ravindranath & Ostwald, 2008). Second, the carbon content is assumed to be 50% of the total biomass, a standard value based on the methodologies provided by the (IPCC, 2006). Finally, the mass of carbon is converted to the mass of carbon dioxide equivalent by multiplying by the ratio of their molecular weights (44/12).

The second service quantified is localized cooling. To achieve this, we developed a novel framework that overcomes the susceptibility of mean-based analyses to thermal outliers. Using Land Surface Temperature (LST) data from Landsat 8/9, the system analyses a 250-meter buffer around each tree to establish a localized thermal baseline. We define two primary metrics. The first, *Maximum Cooling Efficacy* ($C_{\text{eff}}$), represents the tree's performance in mitigating peak surface heat:

$$C_{\text{eff},i,t} = P_{90}(S_i) - LST_{i,t} \quad (2)$$

where $LST_{i,t}$ is the surface temperature of the tree's pixel and $P_{90}(S_i)$ is the 90th percentile LST of non-vegetated pixels within the surrounding buffer $S_i$. The second, *Ambient Heat Relief* ($H_{\text{relief}}$), quantifies the additional cooling benefit a tree provides over other cool, non-living elements:

$$H_{\text{relief},i,t} = P_{10}(S_i) - LST_{i,t} \quad (3)$$

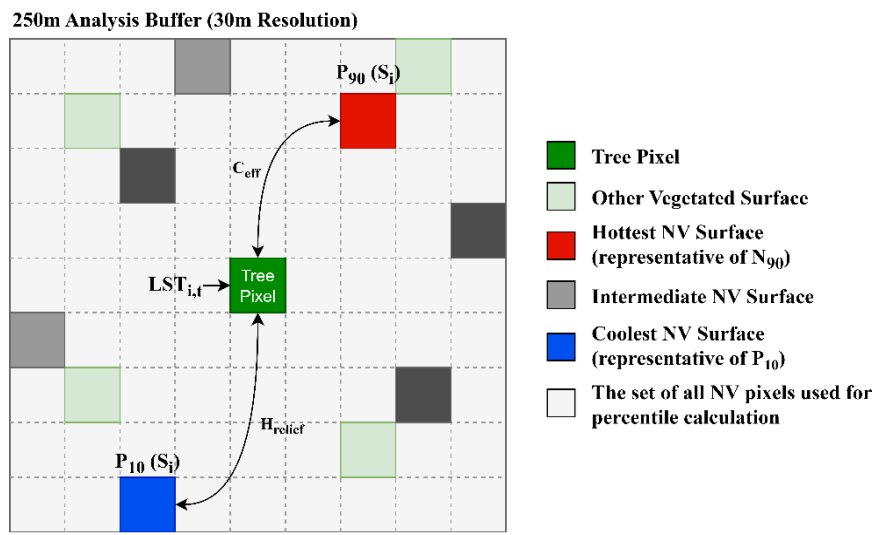

**Figure 4**

*Percentile-based spatial model for tree cooling metrics.*

*Note.* Each tree pixel is analysed within a 250 m buffer of non-vegetated (NV) pixels. The 90th and 10th percentile NV-LST values define the baselines for computing Cooling Efficacy ($C_{\text{eff}}$) and Heat Relief ($H_{\text{relief}}$) enabling localized thermal impact assessment.

To translate these individual performance metrics into generalizable knowledge, we developed an empirical classification system to define Tree Archetypes. This process consists of two steps: (1) Species-Specific Quantile Binning, where tree height, girth, and canopy diameter are independently binned into quartiles (Q1-Q4) for each species; and (2) Archetype Definition, where a composite key is generated by concatenating each tree's quartile assignments. The most statistically frequent combinations are designated as the primary archetypes for that species, yielding semantically rich classes such as "Tamarindus indica - Height:Q4, Girth:Q4, Canopy:Q3".

These analytics are operationalized through a suite of prescriptive, empirically-grounded decision-support tools, including a High-Fidelity Simulation of Planting Potential and a Predictive Model for Thermal Mitigation.

**Module 3: The Eco-Routing Engine for Sustainable Mobility**

The final module operationalizes the outputs of the analytics engine to transform individual mobility by reformulating route selection as a multi-criteria decision analysis (MCDA) problem. The foundational metric is a pre-calculated *Static Environmental Quality* (SEQ) score for each urban street segment. It is defined as a weighted sum of the quantile-transformed values for key environmental attributes:

$$\text{SEQ} = \sum_i w_i \cdot N_q(A_i) \text{ for } i \in \{\text{canopy, CO}_2, \text{biodiversity}\} \quad (4)$$

where $A_i$ is the raw aggregate value for an attribute, $w_i$ is its corresponding weight, and $N_q$ is a Quantile Transformation function. The scoring weights are configurable for different contexts (Refer Appendix A). This non-parametric normalization was deliberately selected to ensure robustness against the highly skewed distributions characteristic of urban ecological data, as shown in Figure 5.

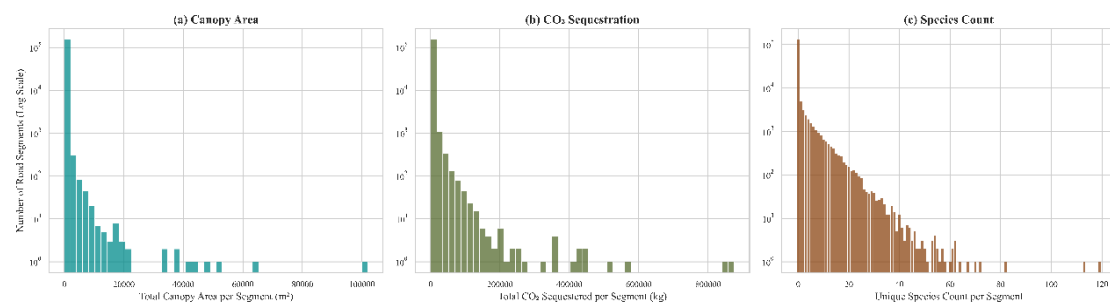

**Figure 5**

*Distribution of Raw Environmental Attributes*

*Note.* The three-panel histogram shows the highly skewed, non-normal distributions for (a) Canopy Area, (b) CO₂ Sequestration, and (c) Species Count per road segment. This illustrates the necessity of non-parametric normalization for calculating the SEQ score.

This SEQ score is integrated into two distinct objective functions. For utilitarian vehicular navigation, the objective is to minimize a Dynamic Holistic Cost (DHC), which treats the SEQ score as a negative cost or an "environmental utility reward." A critical component of the DHC is the predicted vehicular emissions, which are a dynamic function of real-time traffic-adjusted velocity (v). We employ a heuristic polynomial model, a form consistent with established transport emissions research (e.g., Panis et al., 2006), where the emissions factor $E_f$ is defined as:

$$E_f(v) = k_1 + \frac{k_2}{v} + k_3 v^2 \quad (5)$$

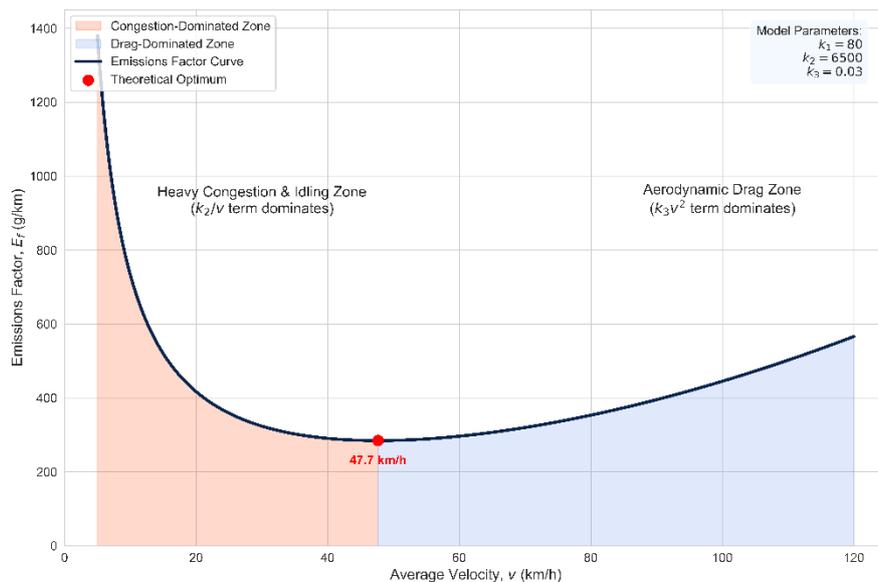

**Figure 6**

*Heuristic Vehicular Emissions Model*

*Note.* The 2D line graph plots the Emissions Factor ($E_f$) in g/km against the Average Velocity (v) in km/h. The U-shaped curve shows the dominance of the congestion and idling term ($k_2/v$) at low speeds and the aerodynamic drag term ($k_3 v^2$) at high speeds.

For recreational activities, the objective shifts to maximizing a *Serenity Score*, which prioritizes tangible human experience. For walking and wellness-focused routes, the serenity score excludes carbon metrics and emphasizes canopy density and biodiversity, using a 0.7 to 0.3 weighting scheme (Refer to Appendix A). Finally, the framework is extended by two algorithms that address non-traditional routing problems: the Optimal Turnaround Algorithm for generating closed-loop, budget-constrained routes and a Resilient Routing Framework that ensures system robustness on maps with heterogeneous data quality. In cases where the foot profile fails due to incomplete pedestrian paths in GraphHopper, the system seamlessly falls back to the car profile, recalculating duration estimates to match walking conditions. (Refer to Appendix A)

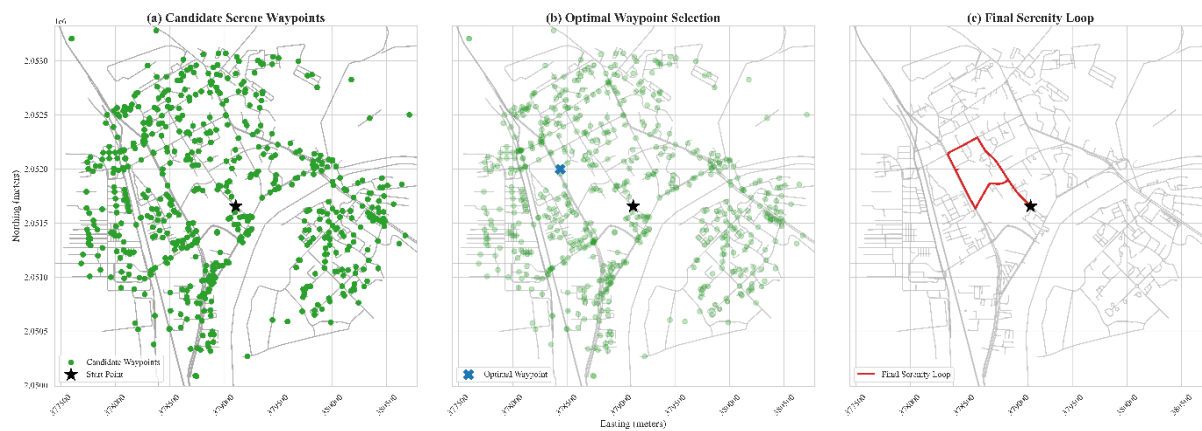

**Figure 7**

*Optimal Turnaround Algorithm Visualization*

*Note*. Visualization of the Serenity Loop algorithm for a 30-minute walk. (a) Candidate waypoints are generated from the centroids of the top 40% of serene road segments within a search radius of the user's start point. (b) An optimal waypoint is selected from the candidates that creates a round-trip route best matching the target distance of 2.5 km. (c) The final out-and-back loop is routed from the start to the optimal waypoint, forming the complete, personalized recreational path.

## Results: A Case Study from Pune, India

To demonstrate the mechanics and potential of the integrated framework, we present a proof-of-concept application of our methodology to the Pune city dataset. It is important to note that the following results constitute a methodological demonstration and a simulated 'case study' rather than an empirical evaluation with end-users, which remains a crucial next step.

**Characterizing the Urban Forest via Tree Archetypes**

The Tree Archetype classification yielded actionable insights for urban planners. Analysis of the Pune Tree Census revealed distinct, statistically common growth forms for key native species. For example, for the species Ficus religiosa, our analysis identified four primary archetypes. As shown in Table 1, the archetype characterized by a tall stature and a wide canopy demonstrated significantly higher cooling efficacy ($C_{eff} = 13.85°C$), making it a prime candidate for targeted planting in heat-stressed urban corridors. This data-driven classification moves beyond generic species recommendations to identify specific, high-performing structural forms for climate-adaptive planting.

| Description (Height, Girth, Canopy) | ($C_{eff}$, °C) | ($H_{relief}$, °C) |
|---|---|---|
| Medium-Tall, Medium-Slender Trunk, Narrow | 13.85 | 5.46 |
| Tall, Thick Trunk, Wide | 13.11 | 4.71 |
| Medium-Tall, Medium-Thick Trunk, Medium-Narrow | 12.75 | 4.28 |
| Various Other Sizes | 11.97 | 3.57 |
| Short, Slender Trunk, Narrow | 11.83 | 3.44 |

**Table 1**

*Sample Tree Archetype Performance for Ficus religiosa in Pune*

*Note*. The table summarizes the performance metrics for different archetypes of Ficus religiosa. $C_{eff}$ refers to the Maximum Cooling Efficacy, and $H_{relief}$ refers to the Ambient Heat Relief.

**Simulating Reforestation Impacts on the Urban Microclimate**

The Interactive Analytics Dashboard allows planners to translate these archetypal insights into spatially explicit scenarios. We simulated a reforestation intervention at a predefined user-made polygon. Using the dashboard's hexagonal packing algorithm, we simulated the planting of 1375 specimens of the top-performing Ficus religiosa archetype. As visualized in Figure 8, the predictive LST model forecasted a significant thermal impact. The simulation predicted a mean Land Surface Temperature (LST) depression of 9.66°C across the park area during peak summer conditions, demonstrating the tool's utility for pre-assessing the benefits of urban greening projects.

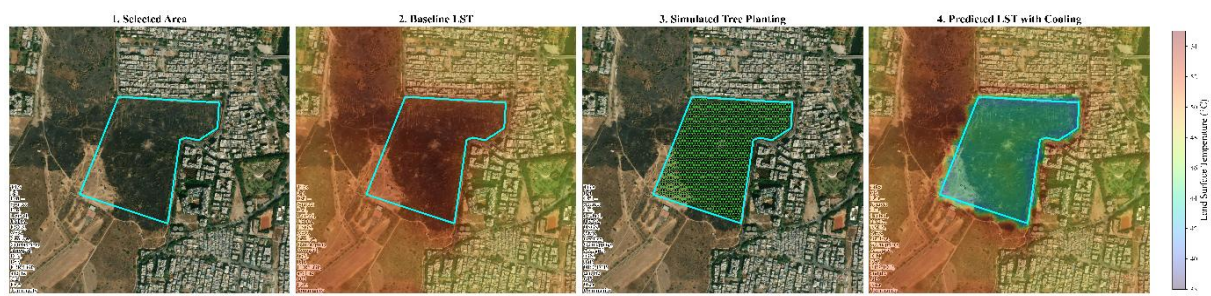

Predictive model showing the cooling effect of a simulated planting of 1375 Guazuma ulmifolia trees.

**Figure 8**

*Visual Abstract of the Predictive LST Model*

*Note*. The four-panel figure illustrates the outcome of a simulated planting intervention, displaying the area's baseline LST and the predicted cooling effect following the simulated planting of high-performance trees.

**Conventional vs. Eco-Friendly Routing: A Comparative Analysis**

The Eco-Routing Engine's primary innovation is its ability to move beyond single-metric optimization by calculating a Dynamic Holistic Cost (DHC) for each potential route.

This cost function, detailed in our methodology, strategically balances travel time, vehicular emissions, and the positive environmental amenities (our Static Environmental Quality score) based on a predefined weighting scheme. The engine's recommendation is therefore not simply the "greenest" or the "fastest" route, but the one that offers the optimal balance between efficiency and ecology.

The comparative analysis in Figure 9 demonstrates a powerful 'win-win' outcome identified by this model. The recommended eco-route is superior across all metrics: it is faster, shorter, produces fewer emissions, and offers a significantly richer ecological experience. This scenario, while not always the case, highlights a key insight: urban infrastructure that prioritizes green corridors can also be highly efficient, challenging the conventional wisdom that sustainability must come at the cost of time.

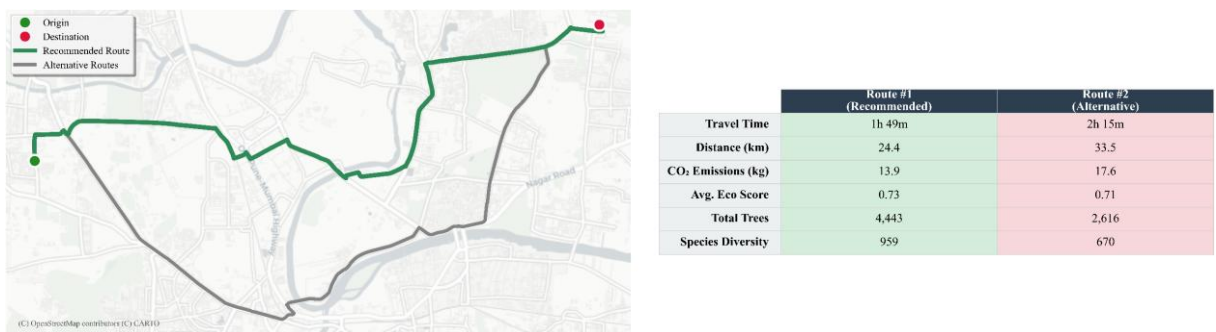

**Figure 9**

*Route Comparison Highlighting Ecological Benefits*

*Note*. The map and accompanying table compare a conventional route with an eco-route, highlighting key metrics such as travel time, distance, CO₂ emissions, and exposure to urban canopy.

## Discussion

### Methodological Contributions and Implications

The primary contribution of this research is methodological. Our introduction of formal, percentile-based cooling metrics ($C_{eff}$, $H_{relief}$) and data-driven "Tree Archetypes" represents a significant improvement for urban planning. This approach replaces generic recommendations (e.g., "plant more trees") with empirically-grounded, context-specific guidance (e.g., "prioritize the planting of Ficus religiosa Archetype 1 in these specific heat-stressed corridors"). The development of client-side simulation tools further empowers planners to pre-assess the impact of greening interventions with high fidelity. Similarly, the multi-criteria routing engine offers a paradigm shift for sustainable mobility by incorporating a positive ecosystem service score (the SEQ). This moves beyond simply minimizing negative externalities, such as emissions, and towards actively maximizing exposure to positive environmental amenities, providing a more holistic definition of a "good" route.

**Empowering Communities Through Open Data**

Beyond technical innovation, our framework provides a direct response to the book's central theme: how open data can empower citizens and democratize urban planning. As Science and Technology Studies (STS) scholars have noted, technologies are not neutral; they co-produce social and political orders (Jasanoff, 2004). In this context, our framework can be seen as an intervention: a tool designed to foster a more inclusive and transparent urban governance. It provides an alternative to centralized, top-down data platforms, which can reinforce existing power structures (van Dijck et al., 2018). By operationalizing the concept of "citizens as sensors" (Goodchild, 2007), our model of smartphone-based crowdsourcing directly addresses the scale mismatch that plagues traditional urban governance. It not only decentralizes datasets that are often inaccessible but also fills critical spatial and thematic gaps in environmental information. By providing transparent, data-driven tools to both planners and citizens, this framework can empower communities to advocate for urban greening initiatives and counteract the commodification trends of neoliberal urbanism, which often prioritize commercial development over public environmental goods. In this context, open data is not merely a repository of information but a catalyst for sustainable urban and transport systems that are inclusive, transparent, and resilient. However, we must critically acknowledge that the promise of such Civic Tech is often unevenly distributed. Tools built on smartphone technology risk exacerbating the 'digital divide,' potentially excluding marginalized communities who lack access to such devices or the requisite digital literacy. Therefore, the successful social integration of this framework would require parallel initiatives focused on digital inclusion, such as public access terminals or community-led data collection workshops, to ensure that the platform empowers all citizens, not just a select few. While our current DHC model uses a fixed weighting scheme designed for an optimal balance, a key area for future research is exploring dynamic weighting based on user preferences. This would allow a user to explicitly define their personal 'acceptance threshold' for travel time, enabling them to choose between a 'balanced' recommendation (our current model) or a 'maximally green' route, even at a higher time cost.

**Limitations and Future Research Directions**

While our framework presents a robust and scalable model, it is essential to acknowledge its inherent limitations, which themselves define clear avenues for future research.

First, and most importantly, the framework is presented here as a proof-of-concept without empirical validation from a user study. A real-world pilot study is the decisive next step to assess the platform's practical usability, the behavioural impacts of the eco-routing engine, and the social dynamics of citizen-led data collection. Such an evaluation, gathering both quantitative analytics and qualitative user feedback, is required to test the real-world effectiveness of our citizen-centred approach.

Second, the methodology is predicated on Land Surface Temperature (LST) derived from thermal infrared remote sensing. LST is a crucial driver of urban heat but is not a direct proxy for near-surface air temperature ($T_a$) or, more importantly, human-perceived thermal comfort (e.g., Mean Radiant Temperature). Second, as discussed in the methodology, the 2D photogrammetric model's validity is fundamentally predicated on the camera being held orthogonally to the measurement plane. We have incorporated features to mitigate common field errors, but significant camera tilt remains a primary source of potential perspective distortion. This directly informs a compelling agenda for future work. A critical next step is to address non-orthogonality by incorporating a Structure from Motion (SfM) pipeline. By allowing a user to capture a short video of a tree, we could generate a true 3D point cloud using tools like COLMAP. This would resolve distortion issues and enable a far more accurate and robust dendrometry analysis. Third, our eco-routing engine does not yet incorporate topography, a significant factor for both vehicular emissions and the feasibility of active mobility. Finally, the system's current implementation provides metric outputs without an associated measure of uncertainty or confidence, a critical component for scientific data collection and building user trust.

**Conclusion**

This chapter has demonstrated a viable, end-to-end methodology for transforming raw urban data into prescriptive analytics for climate adaptation. By synergistically combining citizen science, open data, and advanced analytics, our framework creates a functional feedback loop connecting citizen-led measurement to urban-scale analysis and individual-level action. Our work contributes a replicable, scalable, and participatory model for urban climate action. It shows that by effectively leveraging open data and centring citizens in the process, cities can foster innovative solutions that transcend traditional planning paradigms and create a lasting positive impact on urban environments.

**APPENDIX A**

Two distinct composite scores were used: (1) the Static Environmental Quality (SEQ) score, applied for vehicular routing and carbon-aware planning; and (2) the Serenity Score, used for walking or recreational route suggestions.

The SEQ score combines three normalized environmental indicators associated with road segments: (a) Canopy coverage derived from the sum of tree canopy areas within a 10-meter buffer; (b) Total $CO_2$ sequestered by trees along the segment, based on above-ground biomass calculations; and (c) Tree species richness. These indicators are normalized using a quantile transformation and combined using the following weighted formula:

$$SEQ = 0.5 \times Canopy\ Score + 0.3 \times CO_2\ Score + 0.2 \times Biodiversity\ Score$$

The canopy area was most spatially consistent and visibly impactful; carbon sequestration values varied significantly by species and size; biodiversity was unevenly distributed but valuable for urban cooling and aesthetic purposes. This weighted scheme is modular and can be recalibrated for other cities or datasets, depending on local priorities, e.g., prioritizing biodiversity in ecological corridors, or increasing carbon emphasis in mitigation planning.

The Serenity Score was designed for pedestrian or recreational use cases. It omits carbon metrics to focus on experiential quality, and is defined as:

$$Serenity\ Score = 0.7 \times Canopy\ Score + 0.3 \times Biodiversity\ Score$$

This score favors streets with dense, shady tree coverage and diverse species, enhancing thermal comfort and visual appeal for walkers. Both component scores are again quantile-normalized to reduce the effect of outliers. Route suggestions aim to create loop paths with maximum serenity value while approximating the user's distance target.

In cases where routing via the $foot$ profile failed due to incomplete pedestrian path data in GraphHopper, the system automatically fell back to the $car$ profile to ensure route availability. To preserve pedestrian relevance, distance- and duration-based estimates were recalculated manually based on typical walking speeds. This fallback mechanism maintained user experience reliability while accommodating data sparsity in the underlying routing infrastructure.